\documentclass{aa}
\usepackage{times}
\usepackage{graphics}
\usepackage{xspace}
\usepackage{epsfig}
\usepackage{jwaabib}
\usepackage{rotating}
\usepackage{dcolumn}

\def\Hb{H$\beta$}

\begin{document}


\title{The weak-line T Tauri star V410\,Tau \\ I. A multi-wavelength study of variability}

\author{B. Stelzer\inst{1}
\and M. Fern\'andez\inst{2}
\and V. M. Costa\inst{3,4}
\and J. F. Gameiro\inst{3,5}
\and K. Grankin\inst{6}
\and A. Henden\inst{7}
\and E. Guenther\inst{8}
\and S. Mohanty\inst{9}
\and E. Flaccomio\inst{1}
\and V. Burwitz\inst{10}
\and R. Jayawardhana\inst{11}
\and P. Predehl\inst{10}
\and R. H. Durisen\inst{12}
}

\institute{
  Osservatorio Astronomico di Palermo, Piazza del Parlamento 1, 
  I-90134 Palermo, Italy \and 
  Instituto de Astrof\'\i sica de Andaluc\'\i a, CSIC, Camino Bajo de Hu\'etor 24,
E-18008 Granada, Spain \and 
  Centre for Astrophysics, University of Porto, Rua das Estrelas, 4150 Porto,
Portugal \and 
  Departamento de Matem\'atica, Instituto Superior de 
  Engenharia do Porto, 4150 Porto, Portugal \and 
  Departamento de Matem\'atica Aplicada, Faculdade de C\'\i encas da Universidade do Porto, 4169 Porto, Portugal \and 
  Ulug Beg Astronomical Institute, Astronomicheskaya 33, 700052 Tashkent,
Uzbekistan \and 
  USRA/USNO Flagstaff Station, P. O. Box 1149, Flagstaff, AZ 86002-1149, USA
\and  
  Th\"uringer Landessternwarte, Karl-Schwarzschild-Observatorium, Sternwarte
5, D-07778 Tautenburg, Germany \and 
  Harvard-Smithsonian Center for Astrophysics, 60 Garden Street, Cambridge, MA 02138, USA \and 
  Max-Planck Institut f\"ur extraterrestrische Physik, Postfach 1312, D-85741 Garching, Germany \and 
  University of Michigan, 953 Dennison Building, Ann Arbor, MI 48109, USA \and 
  Indiana University, 727 E. 3rd Street, Bloomington, IN 47405-7105, USA
}

\offprints{B. Stelzer}
\mail{stelzer@astropa.unipa.it}
\titlerunning{A multi-wavelength study of V410\,Tau}

\date{Received $<$16-06-2003$>$ / Accepted $<$04-09-2003$>$}

\abstract{
We present the results of an intensive coordinated monitoring campaign
in the optical and X-ray wavelength ranges 
of the low-mass, pre-main sequence star V410\,Tau carried out in November 2001. 
The aim of this project was to study the relation between various 
indicators for magnetic activity that probe different 
emitting regions and would allow us to obtain clues on the interplay of the 
different atmospheric layers: optical photometric star spot (rotation) cycle, 
chromospheric H$\alpha$ emission, and coronal X-rays. 
Our optical photometric monitoring has allowed us to measure the time of the 
minimum of the lightcurve with high precision. Joining the result with previous data  
we provide a new estimate for the dominant periodicity of V410\,Tau 
($1.871970 \pm 0.000010$\,d). This updated value removes systematic offsets of 
the time of minimum observed in data taken over the last decade. 
The recurrence of the minimum in the optical lightcurve over such a long
timescale emphasizes the extraordinary stability of the largest spot. 
This is confirmed by radial velocity measurements: 
data from 1993 and 2001 fit almost exactly
onto each other when folded with the new period. 
The combination of the new data from November 2001 with published measurements  
taken during the last decade allows us to examine long-term changes 
in the mean light
level of the photometry of V410\,Tau. A variation on 
the timescale of $5.4$\,yr is suggested.  
Assuming that this behavior is truely cyclic V410\,Tau is the first pre-main 
sequence star on which an activity cycle is detected. 
Two X-ray pointings were carried out with the {\em Chandra} satellite
simultaneously with the optical observations, and centered near 
the maximum and minimum levels of the optical lightcurve. A 
relation of their different count levels to the rotation period of the
dominating spot is not 
confirmed by a third {\em Chandra} observation carried out some months later, 
during another minimum of the 1.87\,d cycle. 
Similarly we find no indications for a correlation of the H$\alpha$ emission 
with the spots' rotational phase. The lack of detected 
rotational modulation in two important activity diagnostics seems to argue
against a direct association of chromospheric and coronal emission with the 
spot distribution.  

}

\maketitle

\section{Introduction}\label{sect:intro}

V410\,Tau is an analog for the young Sun on the pre-main sequence (PMS).  
Due to the lack of strong emission lines and infrared excess
it can be classified as a weak-line T Tauri star (wTTS). This
term defines PMS stars without obvious signs for disk accretion,
while young stars, in which accretion from a circumstellar disk 
is responsible for ultraviolet and infrared excess emission and 
for a moderate to strong emission line spectrum superimposed on the
photospheric spectrum, are called classical T Tauri stars (cTTS). 
Indeed, V410\,Tau shows a small infrared excess which, however, 
is attributed to one or two close companions at sub-arcsecond separation  
(\cite{Ghez93.1}; \cite{Ghez97.1}).  

WTTS represent an evolutionary stage where 
the disk has already dissipated, and therefore their variability is not
related to accretion processes but believed to be a manifestation of    
magnetic activity, similar to that observed in the Sun, but
enhanced by several orders of magnitudes. 
Magnetic activity seems to explain all the optical variability 
phenomena discovered on V410\,Tau, 
comprising a large range of time scales: 
{\it i)} variability of the amplitude of the optical light curve 
on time scales of years;
{\it ii)} changes in the H$\alpha$ profile and intensity recorded
within months;
{\it iii)} repeated modulation of the optical brightness with a period of 
$1.87$\,d;
and {\it iv)} sudden brightenings and/or emission line
enhancements evolving on time scales of hours.  

Items {\it i)} and {\it iii)} point directly at the presence of star spots.
The periodic $1.87$\,d variability of the optical lightcurve of V410\,Tau 
was first
reported by \citey{Rydgren83.1}, and confirmed by subsequent observations
(\cite{Vrba88.1}; \cite{Bouvier89.1}). It is attributed to stellar spots
which produce a periodic rotation pattern.   
From these studies it is shown that 
about $15-45$\,\% of the stellar surface of V410\,Tau should be
covered by spots in order to explain the large amplitudes observed in the
optical bands. The models indicated temperatures for the spot(s) that
are $500$\,K to $1200$\,K cooler than
the photosphere, similar to the solar case (\cite{Wallace96.1}).
Independent evidence confirming the spot hypothesis 
was provided by high-resolution spectroscopic observations: 
large ($\geq$10\%) fractions of the stellar surface at about 1000\,K 
below the photospheric temperature can considerably alter the profile 
of photospheric lines (\cite{Vogt83.1}, \cite{Vogt87.1}). 
The first Doppler images of V410\,Tau were published by 
\citey{Joncour94.1} and \citey{Strassmeier94.1}, 
and were confirmed by \citey{Hatzes95.1} and \citey{Rice96.1}. 
All of them show a large cool spot that
extends over one of the stellar poles and smaller cool features located at low
latitudes. 

Longterm variability of the photometric amplitude could be related
to changes in the number, size and location of spots. 
Longterm periodic changes in photometric lightcurves of a number of solar-analogs 
(\cite{Messina02.1}, \cite{Berdyugina02.1}) 
and more evolved RS\,CVn binaries 
(e.g. \cite{Henry95.1}, \cite{Rodono00.1}, \cite{Olah00.1})
have been interpreted as magnetic activity cycles.  
Among younger PMS stars a lack of dedicated monitoring programs has
long impeded any systematic investigation of this issue. 
First results on longterm photometry including PMS stars 
were reported by \citey{Strassmeier97.1} based on observations 
at automated photoelectric telescopes. 
Thanks to the efforts of \citey{Vrba88.1}, \citey{Herbst89.1}, \citey{Petrov94.1}, 
\citey{Strassmeier97.1}, and \citey{Grankin99.1} 
V410\,Tau has by far the largest data base of photometry among the PMS stars.  
Although smooth changes in the amplitude of the optical light
curve have been reported, no cyclic or predictable behavior has been
found yet. 

The origin of variations in emission lines 
[items {\it ii)} and {\it iv)} of the above list] is sometimes less obvious. 
H$\alpha$ observations of V410\,Tau reveal slow changes both in the
emission line profile and its intensity. There are two kinds of changes: smooth
variations of the line profile and intensity within weeks 
(e.g. \cite{Petrov94.1}), that sometimes correlate with the photometric $1.87$\,d 
period (\cite{Fernandez98.1}) and other, stronger, variations that are
noticed when comparing 
data taken months apart from each other (previous references;
\cite{Hatzes95.1}). 
\citey{Petrov94.1} suggest a chromospheric origin for the narrow
emission peak that is often superimposed on a wide platform 
(see also \cite{Hatzes95.1}, \cite{Fernandez98.1}), 
while the platform or flat extended
emission wings might arise from the circumstellar gas environment.

The different line fluxes between observations taken at different
epochs may be due to flaring. Flares are magnetic reconnection events, and
can be as short as $\sim 1$\,h, such that their evolution is
difficult to trace in non-continuous observations. 
For this reason the number of flares reported for V410\,Tau is
quite limited. Most of these events were observed
spectroscopically (e.g. \cite{Welty95.1}), and some flares were
detected also in photometric monitoring programs 
(\cite{Rydgren83.1}, \cite{Vrba88.1}). 

In addition to these well-established activity phenomena observed in 
the optical, magnetic processes further up in the atmosphere,
i.e. the corona, are expected to give rise to radio and X-ray emission.
Indeed, the radio emission of V410\,Tau is highly variable both in intensity 
and spectral index (\cite{Cohen82.1}, \cite{Bieging84.1}),  
leading to the conclusion 
that it must arise from a non-thermal process, 
probably related to magnetic activity (see \cite{Bieging89.1} and
references therein). 
\citey{Bieging89.1} monitored the radio flux density
at monthly intervals over one year. They failed to
detect strong radio flares, but found evidence for a modulation
with a period that is half of the period reported from optical observations.
X-ray data of V410\,Tau has been presented by 
e.g. \citey{Strom94.1}, \citey{Neuhaeuser95.1}, and \citey{Stelzer01.1}.  
\citey{Costa00.1} has extensively discussed archived 
{\em International Ultra-violet Explorer (IUE)} and {\em ROSAT} observations 
of this star. Despite its obvious variability in the X-ray regime no direct signs for
rotational modulation or X-ray flares have been observed from 
V410\,Tau so far. 

The detailed relations among the various atmospheric 
regions involved in stellar magnetic activity 
(photosphere, chromosphere and corona) remain unexplained
and call for a systematic investigation. 
Towards this end we organized contemporaneous
optical and X-ray observations of V410\,Tau. 
Dedicated photometric as well as intermediate- and high-resolution 
spectroscopic observations in the optical were planned 
for a time interval of $\sim 11$\,d,  
simultaneous to three exposures with the {\em Chandra} X-ray satellite. 
This enabled us to study activity on V410\,Tau 
that evolves on short timescales,
i.e. rotational effects and flares. 
In additional we examined historical photometric data of V410\,Tau to 
examine its long-term variability. 

We will present our results in a series of two papers. 
In the present paper we focus on the examination of the relation 
between emission in various energy bands and the optical rotation cycle, 
and the long-term photometry, 
while in a subsequent paper (Fern\'andez et al., in prep.) 
we will concentrate on the numerous flares
found during our campaign. The present paper is structured as follows.
In Sect.~\ref{sect:obs_and_datared} we present the
layout of our observations. We discuss the evolution of the photometric
$1.87$\,d cycle over the years in Sect.~\ref{sect:vis_variab}. This section 
includes a new determination of this period  
and the ephemeris for the minimum brightness, and the
tentative detection of an activity cycle.  
The relation between spots and various activity diagnostics are examined
in Sect.~\ref{sect:rot_act}. We devote Sect.~\ref{sect:xrays} to the description 
of the X-ray spectrum of V410\,Tau. The results are discussed in 
Sect.~\ref{sect:discussion}. Sect.~\ref{sect:conclusion} presents a brief summary.

\section{Observations and Data Reduction}\label{sect:obs_and_datared}

Simultaneous optical and X-ray observations were planned for V410\,Tau
from Nov 15 to 26, 2001 (UT). 
Optical photometry and spectroscopy were 
carried out from several observatories (see below) while three X-ray
observations were scheduled with {\em Chandra} using the Advanced CCD
Imaging Spectrometer for Spectroscopy (ACIS-S). Our
observing strategy was to maximize the probability of detecting large
amplitude variability in the X-ray observations.  If the X-ray emission
comes from spotted regions the maximum of the X-ray emission should
occur at the minimum of the optical lightcurve, i.e. at phases when the
spot is on the visible hemisphere, and vice versa. Therefore, the {\em
Chandra} observations were scheduled for the time near minimum and
maximum of the known rotation cycle, respectively.

\begin{figure}
\begin{center}
\resizebox{9.5cm}{!}{\includegraphics{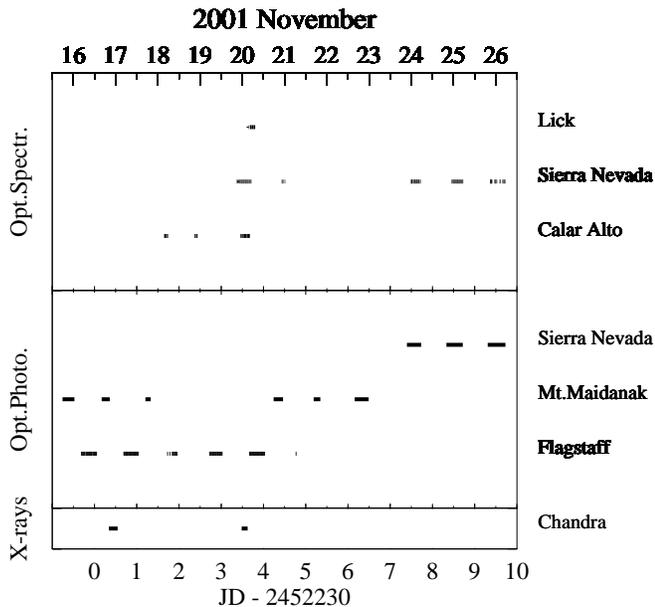}}
\caption{Observing Journal for our monitoring of V410\,Tau in Nov 2001. 
Note, that our campaign includes a third {\em Chandra} observation which was 
performed in March 2002 and is not shown in this diagram.}
\label{fig:obslog}
\end{center}
\end{figure}

Nevertheless, the third {\em Chandra} observation was delayed by
several months because at the anticipated time in November 2001 the
satellite had to be shut down due to high solar activity. Therefore,
optical data is available only for the first two {\em Chandra}
observations.  The complete observing log for the campaign in Nov 2001
is given in Fig.~\ref{fig:obslog}. In the following the individual
observations and the data reduction are described.

%
%

\subsection{Optical photometry}\label{subsect:opt_photo}

\subsubsection{Sierra Nevada}\label{subsubsect:snevada_photo}

The photometric observations carried out at the Observatory of Sierra Nevada
(Granada, Spain) were performed with a Str\"omgren photometer attached to the 90\,cm
telescope. The photometer is equipped with identical six-channel $uvby$
spectrograph photometers for simultaneous measurements in $uvby$ or the narrow
and wide \Hb\/ channels (\cite{Nielsen83.1}). We carried out $uvby$
measurements.  Integration times were 60\,s for V410\,Tau and for its two
comparison stars, HD\,27159 and HD\,283561, as well as for the background
sky. Differential photometry was carried out during all the photometric
nights, except during the last one, in which the absolute calibration was
done.

The error bars of the differential photometry, as estimated from the
difference between the two comparison stars, are $0.005$\,mag in the $u$ band and
less than $0.001$\,mag for the $vby$ bands. Nevertheless, we note that both
comparison stars are about two magnitudes brighter than V410\,Tau,
therefore, slightly larger errors are expected for the target star. 
For the absolute photometry, the residuals of the standard stars (difference
between computed and published values) are less than $0.026$\,mag for the $V$ band
and less than $0.008$, $0.010$ and $0.017$\,mag for the ($b-y$), $m1$ and $c1$ indexes,
respectively.

\subsubsection{Mt.Maidanak}\label{subsubsect:maidanak_photo}

Diaphragm photometry was performed at the Mt. Maidanak
Observatory, Uzbekistan. All observations were obtained with
the 48\,cm telescope equipped with a one-channel pulse-counting 
photometer. We carried out Johnson $UBVR$ measurements 
with a $28^{\prime\prime}$ diaphragm. A detailed description
of the equipment and measuring techniques was given by
\citey{Shevchenko80.1}. The exposure time was typically 40-120\,s,
depending on the band and sky-background brightness.
As a rule, seven standard stars from the list of \citey{Landolt92.1} 
were observed on each night to determine the atmospheric
extinction coefficients. The instrumental photometric system
was reduced to the standard system by the method of \citey{Nikonov76.1}. 
The star BD+28~643 was used as a secondary standard.
The rms error of a single measurement on a moonless night was
$0.032$, $0.007$, $0.005$, and $0.005$\,mag in $U$, $B$, $V$, and $R$, 
respectively.

\subsubsection{USNO Flagstaff}\label{subsubsect:flag_photo}

Observations from the USNO, Flagstaff Station (NOFS, USA) used the
NOFS 1.0m f/7.3 R/C telescope along with a SITe/Tektronix 
$2048 \times 2048$ CCD.
A mask around the CCD limits its usable
area to $1815 \times 1815$ pixels, or about $20 \times 20\,^\prime$. 
By positioning V410\,Tau in one corner of the array 
the largest number of candidate comparison stars could be included. 
Several nights prior to the campaign, plus the photometric nights during the
campaign, were used to calibrate all potential comparison
stars in the field.  $UBVR_{\rm c}I_{\rm c}$ filters were used, along with
a large number of Landolt standard stars (see \cite{Landolt83.1} and
\cite{Landolt92.1}), to
calibrate the comparison stars.  
Five of the potential comparison stars were found to be constant 
during the survey interval, and bright enough to give reasonable 
signal-to-noise (S/N) for differential photometry. 
Differential photometry at $UBVR_{\rm c}I_{\rm c}$ 
was performed extensively on the five nights
Nov 16 through Nov 20 UT, 
plus a few data points on nights prior to and subsequent to this period.

\subsection{Optical spectroscopy}\label{subsect:opt_spect}

\subsubsection{Intermediate-resolution Spectroscopy}\label{subsubsect:snevada_spect}

Intermediate resolution spectroscopy was carried out at the 1.5m telescope on 
the Observatory of Sierra Nevada (Granada, Spain) using the 
spectrograph ALBIREO (see \cite{Sanchez00.1}) in long-slit mode. 
The detector was a EEV8821/T CCD of 1152$\times$770 pixels with a 
22.5\,$\mu$m pixel size. 
Observations were done on two spectral ranges: 4000-5160\,\AA\/ (blue) 
and 5645-6790\,\AA\/ (red). The full width half maximum (FWHM) of
the lines of the calibration lamps were 1.7 \AA\/ and 1.5 \AA\/ for the blue
and red ranges, respectively. 
 Several spectrophotometric standard stars were observed on the three
photometric nights, with the aim of correcting the slope of the
continuum but no flux calibration. 
A set of templates, with spectral types ranging from K2 to K5 were
also observed. 

The data reduction and analysis was done with the {\it longslit} 
package of IRAF
(Image Reduction and Analysis Facility\footnote{IRAF is distributed by the
National Optical Astronomy Observatories, which is operated by the Association
of Universities for Research in Astronomy, Inc. (AURA) under cooperative
agreement with the National Science Foundation.}). The longslit tasks were
required due to the strong distortion of the sky lines along the spatial
direction. 
Individual spectra have a S/N of about 30. 

\subsubsection{High-resolution Spectroscopy}\label{subsubsect:calaralto_spect}

High-resolution spectroscopic observations were performed using
the Fiber Optics Cassegrain Echelle Spectrograph (FOCES) and a
2048$\times$2048 pixel SITe CCD attached to the
2.2m telescope at the Calar Alto Observatory (Almeria, Spain). The spectra
include some seventy orders, covering a range from 4200 \AA\ to 7000
\AA, with a nominal resolving power of $\lambda/\Delta\lambda\,
\approx\,$ 30 000.  Technical details regarding FOCES can be found in
\citey{Pfeiffer98.1}.  

Additionally, high-resolution spectra were taken 
at the University of California's Lick Observatory on Mt. Hamilton 
using the Hamilton echelle spectrograph on the 3m-Coude telescope. 
The instrument yielded $107$ spectral orders spanning a wavelength range of 
$\sim$ 3500 - 10000\,\AA. 
A thinned Ford $2048 \times 2048$ pixel CCD with 15$\mu$m pixel size was used. 
Using an aperture plate above the slit gave a slit 
width of 640$\mu$m, corresponding to $1.2^{\prime\prime}$ projected on the sky, 
and $\sim$ 2\,pixels on the CCD.  This gave a 2-pixel spectral resolution 
of $\sim$ 0.1\,\AA~FWHM (i.e., 2-pixel R $\approx$ 60000).  

All high-resolution spectra have been reduced and extracted 
using the standard IRAF reduction procedures 
(bias subtraction, flat-field division and
optimal extraction of the spectra). Wavelength calibration was done
using spectra of a Th-Ar lamp. Finally, for each order we 
normalized the extracted spectra by means of a polynomial fit to the
observed continuum.

\subsection{X-rays}\label{subsect:obs_chandra}

The {\em Chandra} observations were performed with V410\,Tau on the back-illuminated 
ACIS-S3 CCD. In order to avoid pile-up (the detection of more than one photon as a
single event which
leads to distortion of the spectral shape and underestimate of the count rate) 
the source was placed $4^\prime$ off the aimpoint. This way, for 
the expected X-ray brightness 
of V410\,Tau in the {\em Chandra} energy band, pile-up should be below $10$\,\%. 
However, we performed additional checks on the count rate and the spectrum to verify 
that pile-up is negligible (see Sect.~\ref{sect:xrays}). 
Our data analysis is based on the events level\,1 data provided by the
pipeline processing at the {\em Chandra} X-ray Center (CXC), 
and was carried out using the 
CIAO software package\footnote{CIAO is made available by the CXC and can be downloaded from htp://cxc.harvard.edu/ciao/download-ciao-reg.html} version 2.3.

In the process of converting the level\,1 events file to a level\,2 events file
for each of the observations we performed the following steps: 
We filtered the events file for event grades
(retaining the standard {\em ASCA} grades $0$, $2$, $3$, $4$, and $6$), 
and applied the standard good time interval (GTI) file. 
Events flagged as cosmic ray afterglow 
were retained after inspection of the images revealed that a substantial
number of source photons erroneously carry this flag. 
We also checked the astrometry for any known systematic aspect offsets 
using CIAO software, and performed the necessary corrections by updating the
FITS headers. 

To find the precise X-ray position of V410\,Tau we ran the {\it wavdetect} source 
detection routine (\cite{Freeman02.1}). 
For the source detection we used a binned image of the ACIS-S3 chip 
with $\sim 3^{\prime\prime} \times 3^{\prime\prime}$ pixels. 
The significance threshold
was set to $10^{-6}$, and wavelet scales between $1$ and $8$ were applied. 
After V410\,Tau was located in this way we extracted the 
counts from a circular region of 4$^{\prime\prime}$ radius around the 
position found by {\it wavdetect}. This area includes 
$\approx 97$\,\% of the source photons (for a monochromatic $1.49$\,keV source). 
The background was extracted from an annulus in a 
source-free region surrounding V410\,Tau. 
The observing log for the three {\em Chandra} exposures and the
average ACIS-S count rates of V410\,Tau in the $0.2-8$\,keV energy range 
are given in Table~\ref{tab:x-ray_params}. 
%
%
\begin{table}
\begin{center}
\caption{Observing log for the three {\em Chandra} ACIS-S observations of V410\,Tau. Rotational phases are computed using the new ephemeris given in Table~\ref{tab:ephemeris}. Values refer to the $0.2-8$\,keV energy band.}
\label{tab:x-ray_params}
\begin{tabular}{lrrrr}\\
\hline
Seq.\# & \multicolumn{1}{c}{Start Date} & \multicolumn{1}{c}{Exp. Time} & \multicolumn{1}{c}{$\phi_{\rm rot}$} & \multicolumn{1}{c}{Rate}       \\
       & [d/m/y h:m]                    & \multicolumn{1}{c}{[s]}       &                                      & \multicolumn{1}{c}{[$10^{-3}$\,cps]}   \\
\hline
200130 & 16/11/01 20:11 & $14925.3$ & $0.91-1.00$ & $296 \pm 4$ \\
200190 & 19/11/01 23:37 & $11007.6$ & $0.58-0.65$ & $361 \pm 6$ \\ 
200191 & 07/03/02 06:16 & $17734.2$ & $0.89-0.99$ & $324 \pm 4$ \\
\hline
\end{tabular}
\end{center}
\end{table}

\section{Variability in the Optical}\label{sect:vis_variab}

The rotational period of V410\,Tau was first determined by Rydgren \& Vrba
(1983), who found a value of 1.92\,d from photometric data obtained over 
a 6-day interval. 
\citey{Vrba88.1} combined data taken during five observing seasons between
1981 and 1987, and improved the measurement of the period. 
Only one year later Herbst (1989) noted a linear trend present in the $O-C$ 
(observed minus computed) diagram for minimum light of V410\,Tau based on
the ephemeris by \citey{Vrba88.1} indicating the need for a further
revision. The most recent and most widely used ephemeris for the photometric
minimum of V410\,Tau is the one presented by \citey{Petrov94.1}:   
JD(min)= JD 2446659.4389 + 1.872095($\pm$0.000022)~E. This result 
is based on data from 1986-1992. 

We have used this period and ephemeris to phase-fold the
photometry from November 2001, and found that the minimum is offset 
from phase $\phi=0$ by $\Delta \phi \sim 0.14$ (\cite{Stelzer02.1}). 
A shift of the optical minimum with respect to the ephemeris given by P94
had already been identified by \citey{Grankin99.1} in his investigation
of seasonal lightcurves acquired within 
the last decade. Our measurement continues the monotonic 
trend observed since 1990.
This migration of the minimum could either indicate
a change in the latitude of the spots or the need for a new estimate 
of the period. The structure of the lightcurve of V410\,Tau undergoes
changes over the years (Fig.~\ref{fig:vband_lcs} and Sect.~\ref{subsect:opt_lcs}) 
suggesting that indeed the distribution of surface features is variable in time. 
However, spot models based only on photometric data cannot provide
unique information about the shape and latitude of the spots.

\subsection{Migrating Spots and Differential Rotation}\label{subsect:migrat}

As outlined above, the systematic phase shift of the minimum observed over the last
$\sim 10$\,yrs could be due to a latitudinal migration of the spots 
on the differentially rotating surface. 
The differential rotation of V410\,Tau is known from 
Doppler Imaging, where the differential rotation parameter 
$\kappa = \Delta \Omega / \Omega_{\rm eq}$ ($\Omega_{\rm eq}$ is the rotation
rate at the equator) for an assumed solar-like 
rotation law was found to be $\approx 0.001$ (\cite{Rice96.1}).
This value is much smaller than the solar differential rotation
which is $\approx 0.2$ from equator to pole (\cite{Snodgrass83.1}),
but not untypical for a fast rotating star 
(see e.g. \cite{CollierCameron02.1}).  

The phase shift accumulated over the last decade with respect to the ephemeris
by P94 is $\Delta \phi \approx 0.14$ (see \cite{Stelzer02.1}). This indicates
that $\Delta P / P \approx 6 \times 10^{-5}$.
If we assume a solar rotation law, and that the major spot was centered on the pole 
initially ($\theta_1 = 90^\circ$), this change in period implies that the spot 
should have moved to a latitude of $\theta_2 \approx 75^\circ$ by 2001. 
However, such a large displacement is clearly inconsistent with the relatively 
unchanged shape of the photometric lightcurve. 
Furthermore, Doppler maps obtained in the years 1990 to 1994 seem not to show 
a systematic movement of the main spot.

\subsection{An Update of the Period and Ephemeris}\label{subsect:period} 

In this section we examine the second possibility
for the observed systematic shift of the minimum in the optical lightcurve, 
namely the need for a refinement of the value for the rotational period. 
In the following, whenever we refer to the `rotational period' or 
`rotation cycle' the reader should keep in mind that this is the period 
representative for the main spot that dominates the observed lightcurve. 

\begin{figure}
\begin{center}
\includegraphics[width=9cm, angle=0]{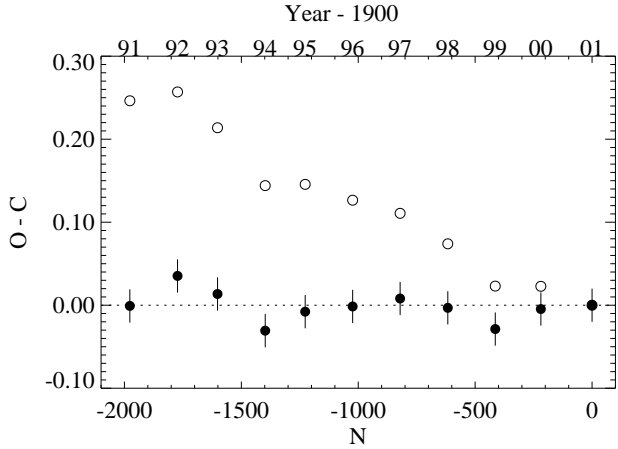}
\caption{$O-C$ diagram for the seasonally averaged times of minimum $T_0$ in the $V$ band lightcurve of V410\,Tau versus rotational cycle number $N$. {\em Open circles} - for the period given by P94, and {\em filled circles} - after adapting the period to minimize the residuals. Data obtained in Nov 2001 were used as reference point.}
\label{fig:o_minus_c}
\end{center}
\end{figure}

For our update of the period we made use of historical data from the T Tauri 
photometry data base.\footnote{The T Tauri photometry data base is compiled
and maintained by W. Herbst and available at http://www.astro.wesleyan.edu/$\sim$bill/}
We split the available $V$ band photometry on a yearly basis, and derived 
the most significant 
period for each of the observing intervals using the string length
method (\cite{Dworetsky83.1}). 
Each seasonal lightcurve was folded with its period to determine the 
time of minimum $T_0$ by fitting a polynomial to the folded lightcurve. 
We used the value for the period given by P94 to compute the number of 
cycles elapsed between each of the seasonal points 
$T_0$ and the time of minimum observed by us in November 2001, 
using this last observation as a reference point. 
Then we computed the $O - C$ diagram for the minima
$T_0$ as a function of cycle number.
The residuals in this diagram can be minimized by modifying the period. 
But the $O - C$ residuals show systematic non-linear trends for data 
acquired before 1990. The large scatter for these earlier years may indicate
that the spots migrated in an irregular way on the surface of V410\,Tau.
For years later than 1990 we observe a monotonic trend in the $O - C$ 
residuals. Therefore we used only observations obtained in 1991 and later
for the final determination of the period.
This way a best fit period of $1.871970(10)$\,d is found 
(see Table~\ref{tab:ephemeris} for the new ephemeris of V410\,Tau),
slightly lower than the earlier determination by P94 which was based on data
from 1986 to 1992.
\begin{table}
\caption{Ephemeris of V410\,Tau: value by P94 and our update.}
\label{tab:ephemeris}
\begin{center}
\begin{tabular}{llr} \hline
Reference & \multicolumn{1}{c}{$T_{\rm min}\,{\rm [JD]}$} & \multicolumn{1}{c}{$
P_{\rm rot}\,{\rm [d]}$} \\
\hline
P94   & $2446659.4389$   & $1.872095 \pm 0.000022$ \\
here  & $2452234.285971$ & $1.871970 \pm 0.000010$ \\
\hline
\end{tabular}
\end{center}
\end{table}
The residuals in the $O - C$ plot for the new best fit period are shown in 
Fig.~\ref{fig:o_minus_c}. 
In the following we use the ephemeris and period newly derived here.
%
%

\subsection{Changes in the spots}\label{subsect:opt_lcs}

Provided the newly derived ephemeris represents the correct period for the
surface structure on V410\,Tau, the minima of phase folded lightcurves
for each observing season should now coincide with $\phi = 0$. 
This is verified in Fig.~\ref{fig:vband_lcs} where the phase plots of the 
yearly averages of the $V$ band lightcurve of V410\,Tau 
since 1981 are shown. 
Photometry from earlier years is available only in graphical form,
and the time information can not be extracted accurately enough for studies
of the $1.87$\,d periodicity (original references \cite{Romano75.1}, \cite{Roessiger81.1}). 

The minima of all years included in our determination of the period, 
i.e. 1990 to 2001, are aligned at the same phase, as expected.  
The minima of the years 1986 and 1989 do show a shift.  
This phase migration seems to be irregular and may indeed indicate 
changes in the spot location or distribution. 
Between 1981 and 1985 the lightcurve of V410\,Tau was characterized
by a completely different, double-peaked structure. 
Data from these years has been extensively
discussed by \citey{Herbst89.1} who showed that a two-spot model provides
a good description of the visible lightcurve. 

Changes in the shape of the lightcurve reflect variations of the structure of 
active regions. In order to quantify these changes  
we plot the time-evolution of amplitude, mean, minimum, and maximum 
of the $V$ band lightcurve in Fig.~\ref{fig:vband_longterm}. 
Over the years the lightcurve of V410\,Tau has systematically become 
more stable and its variability more regular. 

From a simple
harmonic fit to the mean $V$ band magnitude for data obtained after 1990 
(also displayed in Fig.~\ref{fig:vband_longterm}) 
a tentative cycle length of 5.4\,yr is derived. 
Inspection of the photometry in other bands, also available at
the T Tauri photometry data base, shows the same
5.4\,yr-pattern in the $B$ band with a similar amplitude, supporting
the idea that this variation can be attributed to a spot activity cycle. 
The data in the $R$ and $I$ band is scarce and inhomogeneous because of the
use of different filters (Johnson and Cousin). 

However, when extrapolating the 5.4\,yr-periodicity to earlier years 
(dotted line in Fig.~\ref{fig:vband_longterm})  
the data first runs out of phase with the $5.4$\,yr period 
and then turns into completely irregular behavior. 
The evolution of both the shape of the lightcurve and the mean brightness
are suggestive of a transition towards less active behavior. 
%
%
\begin{figure}[h]
\begin{center}
\parbox{8.5cm}{\includegraphics[width=8.cm, angle=0]{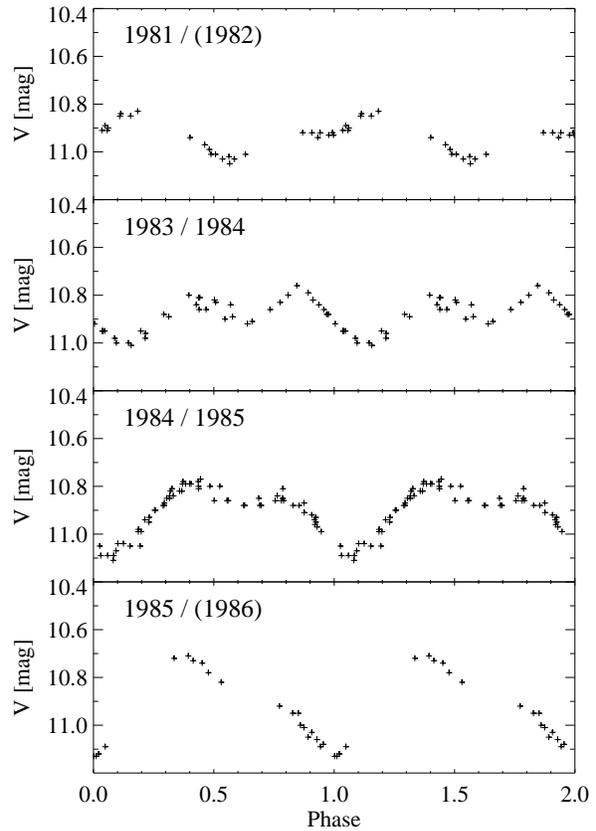}}
\caption{Seasonally averaged $V$ band lightcurves of V410\,Tau from 1981  
to 2001 phase folded with the new ephemeris given in Table~\ref{tab:ephemeris}
(continued on the next page). 
Data are extracted from the data base maintained by W. Herbst.
The data for 2001 was obtained by one of us (KG) 
in the months prior to our coordinated observing campaign. 
} 
\label{fig:vband_lcs}
\end{center}
\end{figure}

\addtocounter{figure}{-1}

\begin{figure*}
\begin{center}
\parbox{8.5cm}{\includegraphics[width=8.cm, angle=0]{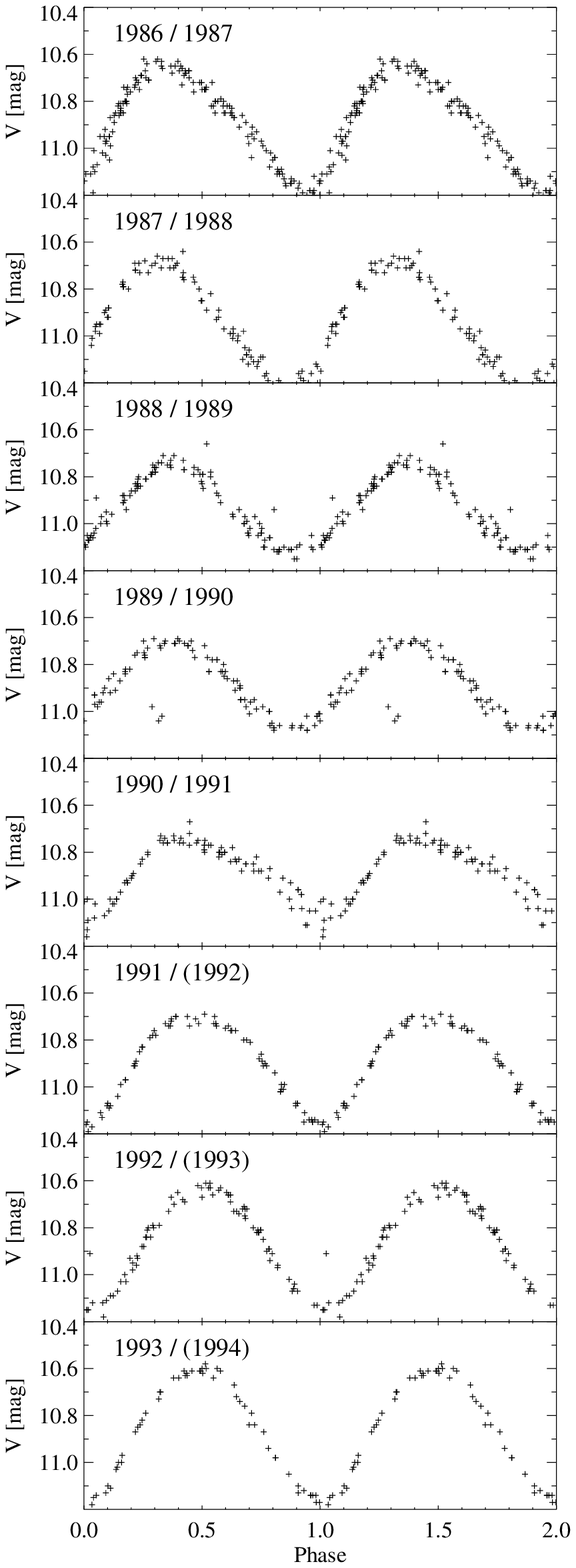}}
\parbox{8.5cm}{\includegraphics[width=8.cm, angle=0]{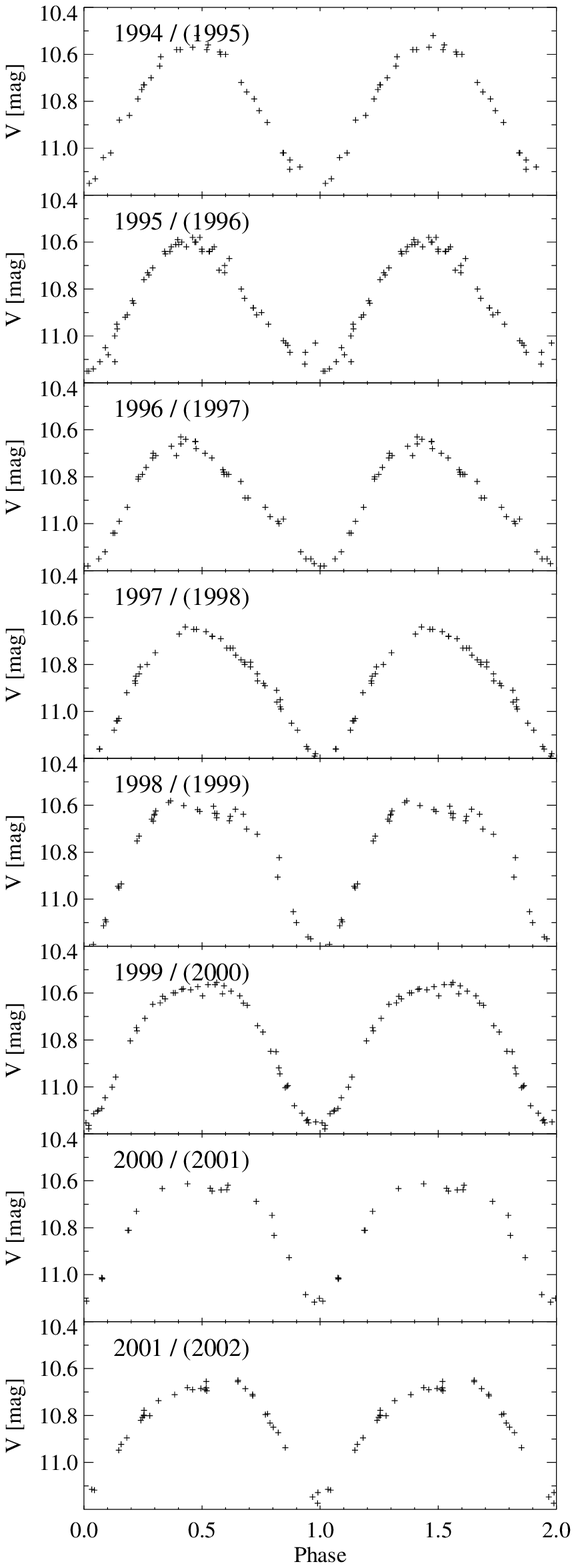}}
\caption{{\em continued}} 
\end{center}
\end{figure*}

%
%
\begin{figure*}
\begin{center}
\includegraphics[width=18.cm, angle=0]{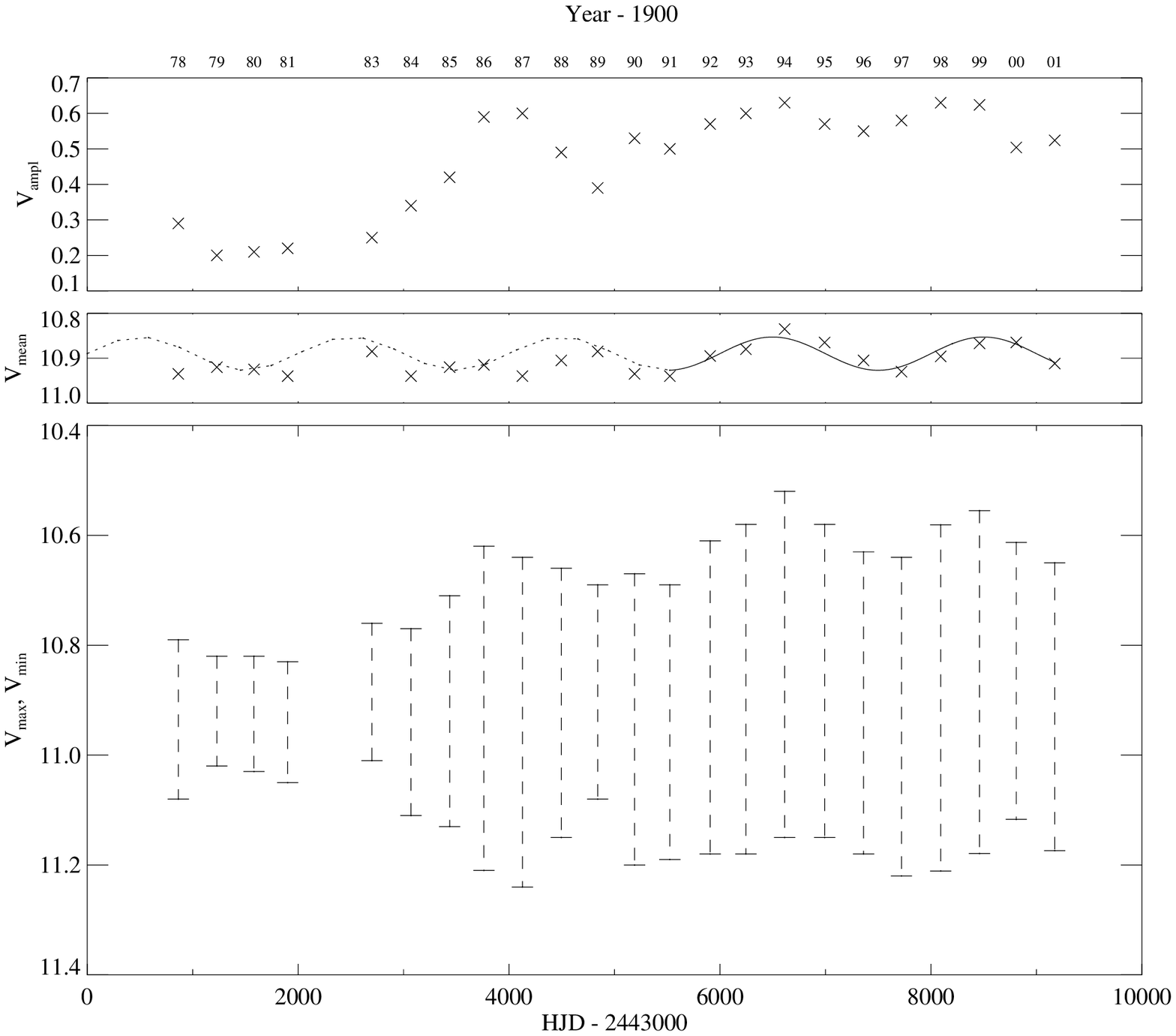}
\caption{Longterm behavior of amplitude, mean, and maximum and minimum of the $V$ band lightcurve of V410\,Tau from 1978 to 2001. 
Data points represent seasonal averages. The numbers on the top y-axis stand for the year in which the observing season started.} 
\label{fig:vband_longterm}
\end{center}
\end{figure*}

\section{Simultaneous X-ray and Optical Data}\label{sect:rot_act}

A major aim of this campaign was to study simultaneous X-ray
and optical observations in order to look for correlations, in particular
those related to the spot rotation cycle. To this end the 
{\em Chandra} observations were purposely scheduled to cover different 
rotational phases of the star. 
If X-ray emission is related to spotted regions the
maximum of the X-ray emission should occur at the minimum of the optical
lightcurve, i.e. at times when the spot is on the visible hemisphere, and 
vice versa. Therefore, to maximize the amplitude of the expected X-ray 
variability we observed near optical maximum and minimum.
In addition our spectroscopic optical monitoring provided information
on the time-evolution of the H$\alpha$ line and the radial velocity (RV). 

\begin{figure*}
\begin{center}
\includegraphics[width=18.5cm, angle=0]{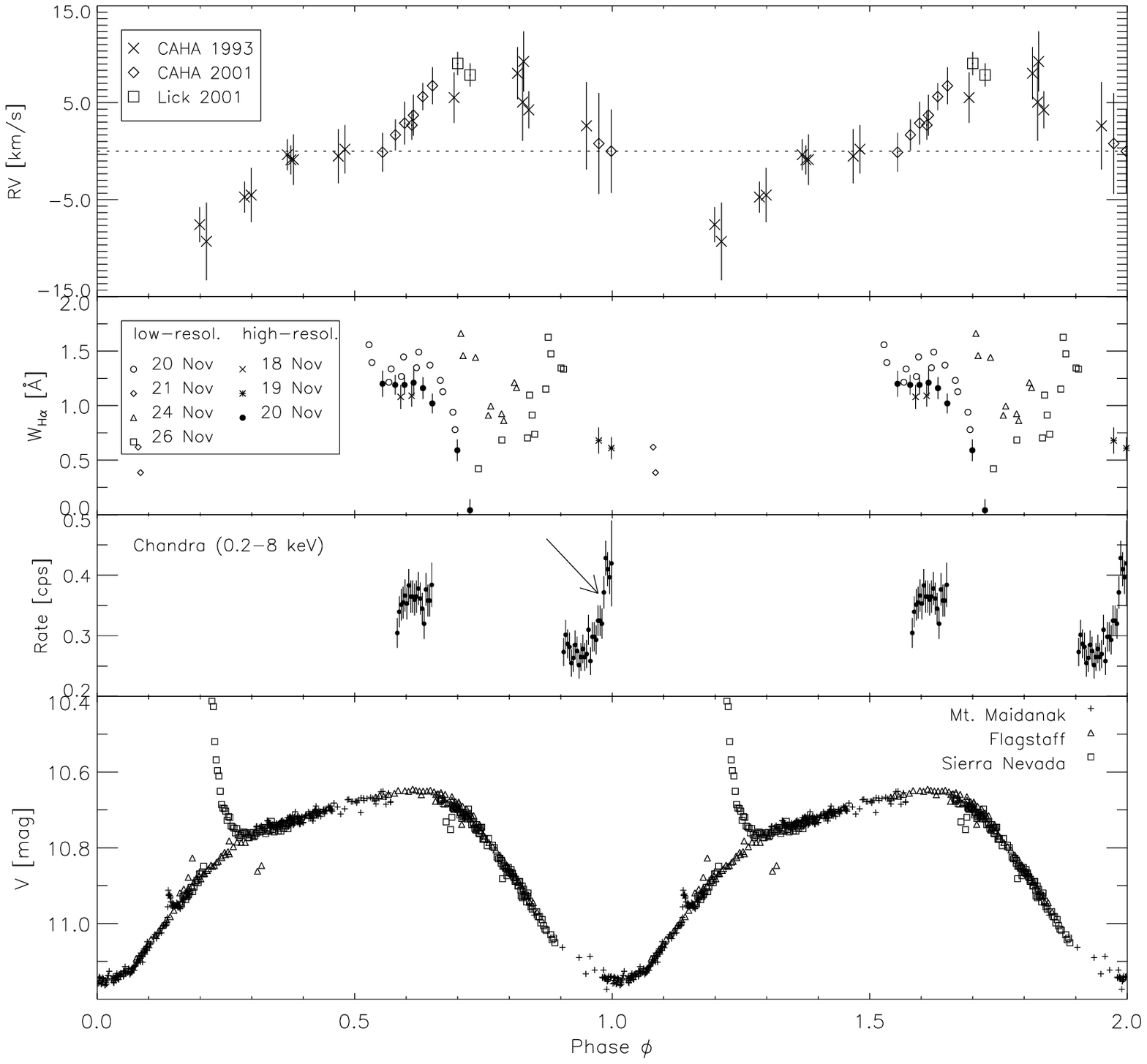}
\caption{Multi-wavelength phase plot for the 1.87\,d cycle of V410\,Tau. 
From top to bottom: Radial velocity, equivalent width of H$\alpha$, 
X-ray count rate, and $V$ band photometry. 
Except for some data points in the radial velocity curve all measurements are 
obtained quasi-simultaneously in November 2001. 
The large flare visible in the $V$ band lightcurve
was also observed spectroscopically, but the equivalent widths of
H$\alpha$ are outside the range displayed in this figure (see
Fern\'andez et al., in prep. for a detailed discussion). 
The fast increase of the X-ray count rate near $\phi = 0$
(marked with an arrow) 
probably indicates the beginning of a flare (see Sect.~\ref{sect:xrays}).}
\label{fig:multil_lcs}
\end{center}
\end{figure*}

The $V$ band lightcurve obtained during our monitoring in November 2001   
is shown in the lowest panel of Fig.~\ref{fig:multil_lcs}
phase-folded with the new period and ephemeris. 
The differential photometry carried out in the Str\"omgren  
system was transformed to the Johnson system with help of the
absolute Str\"omgren photometry done during the last night.  
We point out that the photometric measurements from the three observing sites 
complement each other to provide nearly full coverage of the rotational cycle
despite the relatively short duration of the monitoring (11\,d; 
see Fig.~\ref{fig:obslog}). Photometric data in the $UBRI$ bands was
also obtained, but does not provide new information on the {\em phasing} of the 
$1.87$\,d cycle.
We will use these latter lightcurves for the discussion of flares in 
an accompanying paper (Fern\'andez et al., in prep). Some of these flares
can be seen in the $V$ band lightcurve displayed in Fig.~\ref{fig:multil_lcs}.
However, our aim in this section is to examine activity parameters
for a possible relation to the $1.87$\,d cycle, and short-term random 
processes such as flares are not of interest.  
%
%

\subsection{X-rays}\label{subsect:phase_x-rays}

The mean ACIS count rates listed in Table~\ref{tab:x-ray_params} indicate
that the X-ray emission of V410\,Tau was variable. 
To check the relation with the optical photometry we generated X-ray 
lightcurves, and phase-folded them with the $1.871970$\,d period derived in
Sect.~\ref{subsect:period}. 
The result for the two observations obtained in November 2001 
is shown in Fig.~\ref{fig:multil_lcs}. 
A rapid rise of the X-ray count rate took place near optical minimum in the 
observation from November, probably indicating the onset of a flare. 
This event was not accompanied by simultaneous observations in the optical, 
but just a few hours later two flares were observed
from Mt.Maidanak, hinting at a possible relation between the optical and 
X-ray emission sites. 

A third {\em Chandra} pointing was obtained in March 2002
near optical minimum, i.e. coinciding in phase 
with the first November observation. As this observation is not simultaneous
with the optical lightcurve we do not display it in Fig.~\ref{fig:multil_lcs}. 
The March data does not
show significant time-variability, and its count rate is somewhat higher
than the lowest (quiescent) level measured in the November observation 
taken at the same rotational phase. 
Thus, while the two November pointings 
seem to suggest that the X-ray emission depends on rotational phase, 
this conclusion is not supported by the observation from March.

\subsection{H$\alpha$}\label{subsect:phase_halpha}

The H$\alpha$ line is in the spectral range of both the intermediate-
and the high-resolution spectra obtained within this campaign. 
H$\alpha$ is always weak in emission. We display the equivalent width $W_{\rm H\alpha}$  
measured at different phases of the photometric rotation cycle in the second 
panel of Fig.~\ref{fig:multil_lcs}. 
On Nov 20 both intermediate- and high-resolution spectra were obtained,
and the measured equivalent widths are in good agreement within the uncertainties.
The error bars for the high-resolution spectra are based on the assumption of an
accuracy in the continuum normalization of 10\,\%. The typical error bar for the
values from the low-resolution spectrum are $\sim 30$\,\%.  

On Nov 24 a large flare erupted which we monitored simultaneously 
in spectroscopy and photometry. Note, that the event is not seen in 
the equivalent width curve shown in Fig.~\ref{fig:multil_lcs} because the
values for $W_{\rm H\alpha}$ measured during the flare are higher than the 
displayed plot range.  

Due to poor weather conditions, the phase coverage of the H$\alpha$ data is 
limited. While the equivalent width seems to undergo
(short-term) changes within each night, no clear trend related to the
rotational phase can be found.

\subsection{Radial Velocity and $v \sin{i}$}\label{subsect:phase_rv}

We derived the RV and the projected rotational 
velocity ($v\sin{i}$) from the optical high-resolution spectra 
using a cross-correlation
analysis with several template stars of spectral class K2, K4 and K5.
 
The value of $v\sin{i}$ is obtained by comparing the spectra of V410\,Tau 
within several spectral orders to that of a broadened template. 
Amongst the templates at our disposal, the
K2\,V star HD\,166620 broadened to a velocity of $74\,\pm\,3$\,km/s
gave the best fit. This broadening agrees with the estimates of 
\citey{Vogel81.1} ($76 \pm 10$\,km/s), 
\citey{Hartmann86.1} ($70.9 \pm 9$\,km/s), 
and \citey{Hatzes95.1} ($77 \pm 1$\,km/s). 

We estimated the stellar RV from the data acquired in Nov 2001. 
This was done by averaging measurements that were obtained at 
phases $0.5 < \phi < 0.6$. The four spectra obtained in this phase interval 
correspond to the maximum in the lightcurve, and 
therefore their shape should be less distorted by the spots. 
Using this procedure we found an average RV 
of $17.9 \pm 1.8$\,km/s similar to the stellar RV of V410\,Tau 
given in the literature ($18$\,km/s, \cite{Herbig88.1}). 

The relative shift between photospheric lines was then measured by cross-correlating 
the spectra of V410\,Tau with the spectrum of the template HD\,166620. 
Three different orders have been used, 
with wavelengths centered at $5400$, $6100$ and $6400$~\AA, and a
wavelength coverage of $80$\,\AA\ in each order. 
A parabolic function was used to find the center
and width of the correlation peak, and the errors were
computed from the fitted peak height and the asymmetric noise as
described by \citey{Tonry79.1}. 
The final RV are an average of those obtained with
the three orders, and the accuracies are the combined error bars for each
individual measurement. 

We performed this procedure for each of the 12 
individual high-resolution spectra
of V410\,Tau obtained during our campaign.  
The resulting RV curve is shown in the top panel of 
Fig.~\ref{fig:multil_lcs} as a function of rotational phase, combined with a number
of measurements obtained in 1993 and presented by \citey{Fernandez98.1} added here
to improve the phase coverage.  

From this latter data set we used only the 
spectrum in the range $6680 - 6730$\,\AA. This spectral range is useful 
because strong photospheric lines such as the Li I line ($6708$\,\AA) 
are present. In a first step, each of the $16$ spectra from 1993 
was cross-correlated with the first spectrum of the 1993 series. 
The position of the maximum of the cross-correlation function
gives then for each spectrum the average shift of the photospheric lines 
relative to the first one.
To be able to combine these results with the
RV measurements obtained during 2001 we took the velocity shift at
rotational phase $\phi = 0.5$ as a zero-point. 
At this phase the photometry shows that the star is at its brightest, 
and thus the spot(s) or a large part of it is not visible.  
As a consequence, 
we expect to see the normal stellar photosphere,  
and the velocity shift
should be equal to the stellar RV, i.e. $\Delta V_{\rm rad}=0$.

From Fig.~\ref{fig:multil_lcs} it is readily seen that the RV 
varies from $\approx -10$ to $+10$\,km/s, 
and the phase shift is $\approx 0.25$ when compared with the photometric 
lightcurve. This phase shift can be
understood if the cool spot is responsible for the distortion of
absorption lines (\cite{Vogt87.1}). 
In this scenario, the RV is large when
the spot is near the limb and close to zero when the spot is face-on
($\phi\,\approx\,1.0$) or occulted by the star ($\phi\,\approx\,0.5$).
This result shows how magnetic stellar activity can
affect the spectroscopic search for very low-mass companions orbiting
around PMS stars.

\section{X-ray Properties}\label{sect:xrays}

We extracted an X-ray spectrum for each of the three {\em Chandra} 
pointings. The observation from 
Nov\,16/17 shows a pronounced increase in count rate at the end of the
exposure (discussed above in Sect.~\ref{subsect:phase_x-rays}). 
If this behavior in the X-ray lightcurve is due to a flare,  
it is expected that the spectrum changes as a consequence of coronal heating. 
To unveil any eventual spectral variability in the data 
we split the Nov\,16/17 observation into a quiescent and flaring part, 
separating the data at JD\,2452230.4870 (marked with an arrow in 
Fig.~\ref{fig:multil_lcs}). 
No systematic trend is seen in the temporal behavior of the
other two {\em Chandra} pointings. Therefore, we think of both as 
representing the quiescent, i.e. non-flaring, state of V410\,Tau.

As outlined in Sect.~\ref{subsect:obs_chandra} we avoided excessive photon
pile-up by placing V410\,Tau at an off-axis angle of $\sim 4^\prime$. 
To check whether any remaining pile-up affects the spectrum the source extraction 
area was varied: Next to the spectrum from a $4^{\prime\prime}$
radius centered on the {\it wavdetect} position of V410\,Tau 
we extracted the spectrum from an annulus where we excluded the inner source region. 
In the outer portions of the PSF the count rate is smaller and no pile-up
is expected. Visual inspection does not show any distortion of the full-source 
spectrum with respect to the one extracted from a (with certainty pile-up free) 
annulus. The supposition that pile-up is negligible was confirmed  
when we fitted both spectra with the same model and found no significant 
differences in the spectral parameters. 
In the following we describe the details about our spectral model. 

Spectral fitting was done in the XSPEC environment (version 11.2.0).  
To make up for the continuous degradation of the ACIS
quantum efficiency we applied the {\it acisabs} model to the 
auxiliary response file (arf) before loading the data into XSPEC. 

We applied the MEKAL model (\cite{Mewe85.1}) for thermal emission from an 
optically thin, hot plasma plus a photo-absorption term. 
Comparison of the three quiescent spectra
(first part of Nov\,16/17, Nov\,19/20, and Mar\,7)
shows that their spectral shape is very similar. Separate fitting led to
indistinguishable parameters. Therefore, we present here the result of 
a joint modelling of all three quiescent spectra which improves 
the statistics. 
To represent the observed small offsets in the flux of the 
three spectra we allowed for an independent normalization constant.

\begin{figure}
\begin{center}
\resizebox{9cm}{!}{\includegraphics{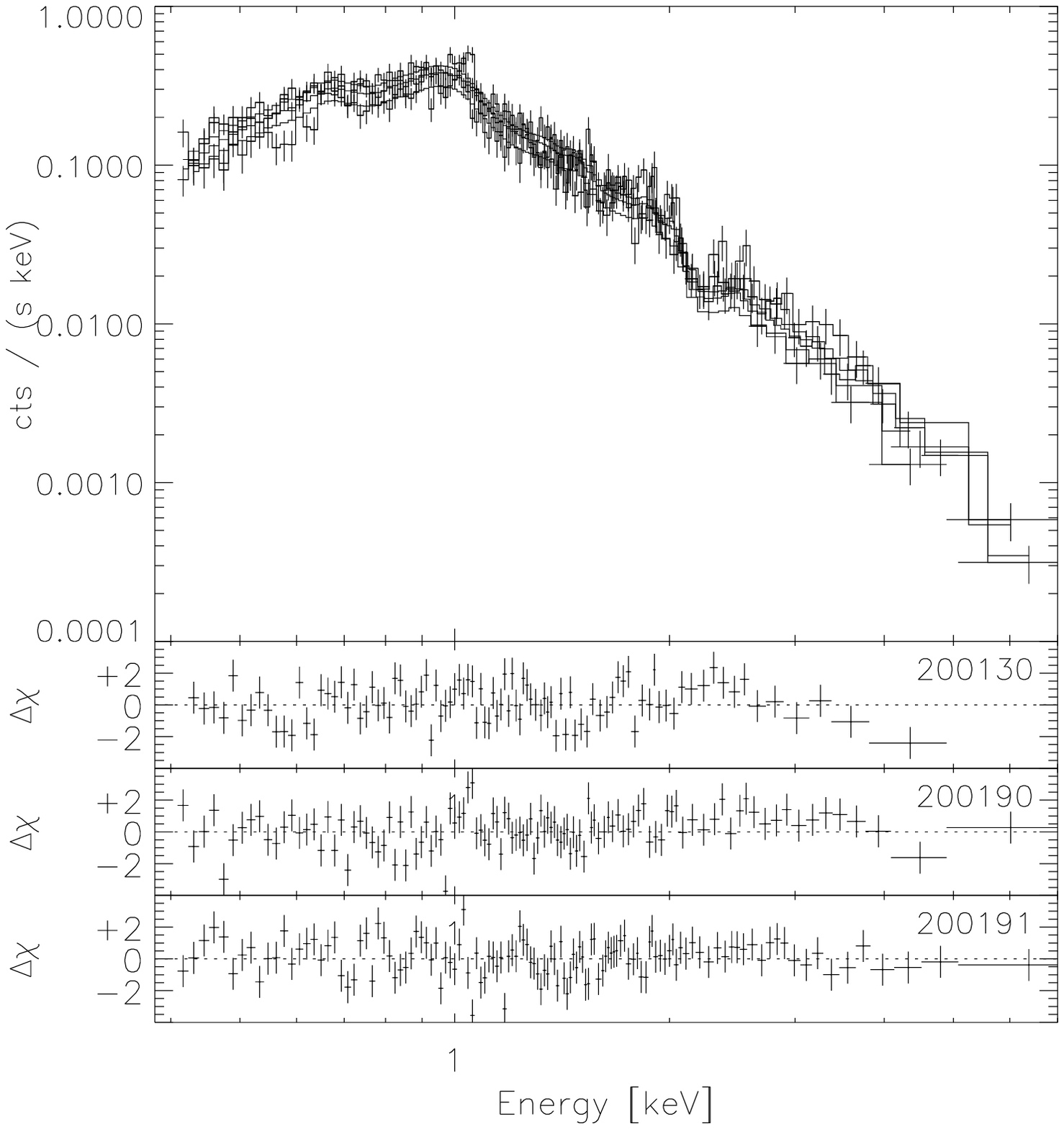}}
\caption{Quiescent X-ray spectrum of V410\,Tau during three {\em Chandra} exposures. A flare has been eliminated from Obs. Seq.\#~$200130$. The data are overlaid by the best fit model from Table~\protect\ref{tab:x-ray_spectra}. The residuals for each observations are shown in separate panels.}
\label{fig:x-ray_spectra}
\end{center}
\end{figure}

An acceptable fit ($\chi^2_{\rm red} \sim 1$ and flat residual) 
requires a minimum of three MEKAL components, and free elemental abundances.
The individual spectra, the best fit model and residuals are displayed in 
Fig.~\ref{fig:x-ray_spectra}. 
In Table~\ref{tab:x-ray_spectra} we summarize the spectral parameters
observed during the quiescent state.
%
%
%
%
\begin{table}
\begin{center}
\caption{X-ray spectral parameters of V410\,Tau derived from a joint fit to the three ACIS observations that represent the stars' quiescent state. A constant normalization factor was applied to allow for the observed difference in X-ray brightness between the three data sets: $N_{\rm 200190} = 1$, $N_{\rm 200191} = 0.91 \pm 0.04$, $N_{\rm 200130} = 0.74 \pm 0.04$. Errors are 90\,\% confidence levels.}
\label{tab:x-ray_spectra}
\begin{tabular}{llr}\\ \noalign{\smallskip} \hline\noalign{\smallskip}
Parameter     & Unit   & Fit result \\ \noalign{\smallskip} \hline\noalign{\smallskip}
$N_{\rm H}$   & [${\rm 10^{21} cm^{-3}}$] & $0.97^{+0.37}_{-0.32}$ \\ \noalign{\smallskip} \hline \noalign{\smallskip}
$kT_1$        & [keV]  & $0.24^{+0.03}_{-0.02}$ \\ \noalign{\smallskip} \hline \noalign{\smallskip}
$kT_2$        & [keV]  & $0.93^{+0.09}_{-0.07}$ \\ \noalign{\smallskip} \hline \noalign{\smallskip}
$kT_3$        & [keV]  & $2.18^{+0.43}_{-0.28}$ \\ \noalign{\smallskip} \hline \noalign{\smallskip}
$EM_1$        & [${\rm 10^{53}\,cm^{-3}}$] & $3.02^{+1.01}_{-1.24}$ \\ \noalign{\smallskip} \hline \noalign{\smallskip}
$EM_2$        & [${\rm 10^{53}\,cm^{-3}}$] & $2.23^{+1.26}_{-0.80}$ \\ \noalign{\smallskip} \hline \noalign{\smallskip}
$EM_3$        & [${\rm 10^{53}\,cm^{-3}}$] & $2.13^{+0.45}_{-0.77}$ \\ \noalign{\smallskip} \hline \noalign{\smallskip}
$Z$           & [$Z_\odot$]       & $0.20^{+0.08}_{-0.06}$ \\ \noalign{\smallskip} \hline \noalign{\smallskip}
\multicolumn{2}{c}{$\chi^2_{\rm red}$ (dof)}  & $1.30$ ($330$) \\ \noalign{\smallskip} \hline \noalign{\smallskip}
\end{tabular}
\end{center}
\end{table}

\begin{figure}
\begin{center}
\resizebox{9cm}{!}{\includegraphics{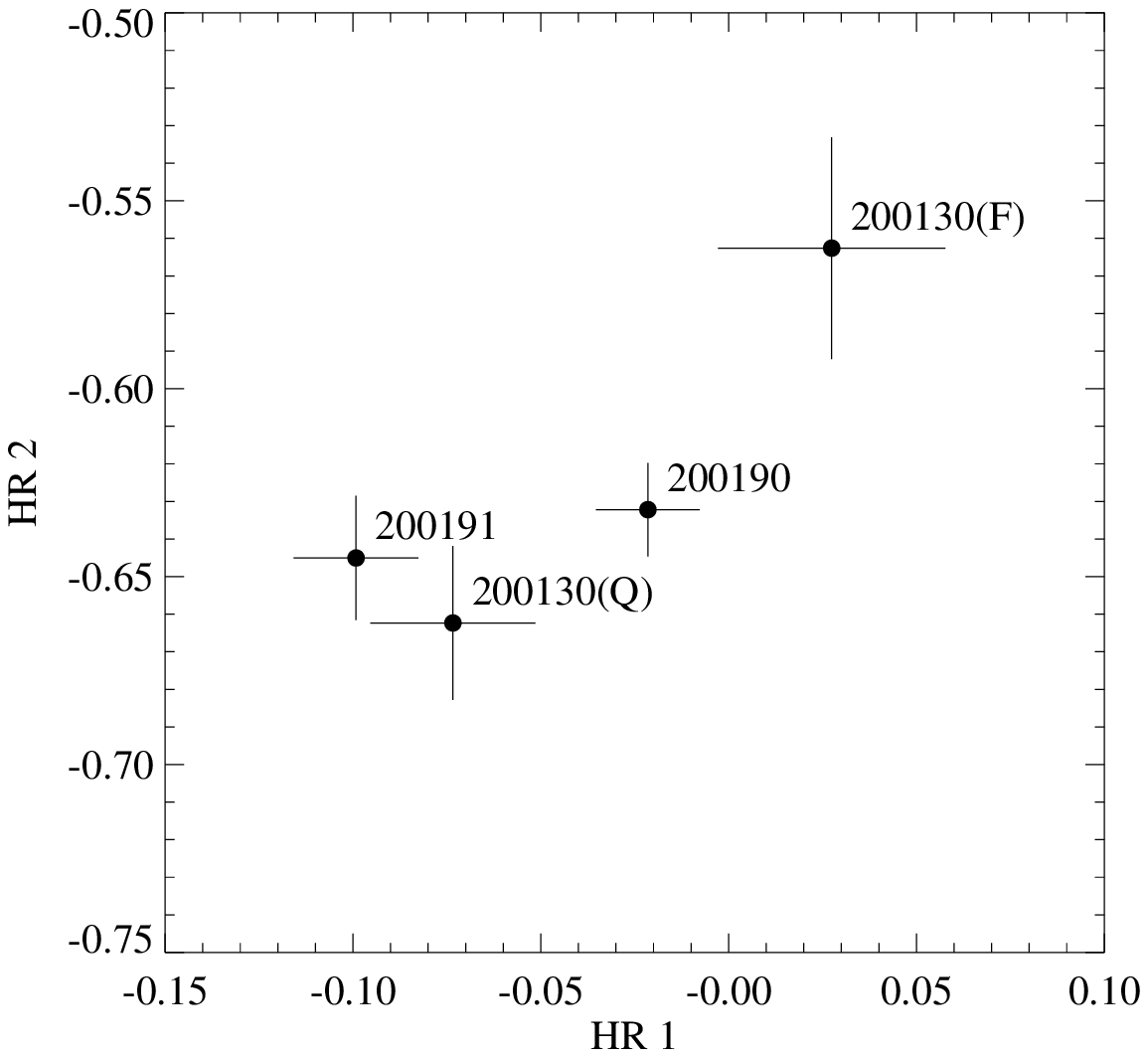}}
\caption{ACIS hardness ratios for V410\,Tau. Flare and quiescent state for observation 200130 are labeled with `F' and `Q', respectively. Error bars denote 1\,$\sigma$ uncertainties.}
\label{fig:hrs}
\end{center}
\end{figure}

To examine whether the rise in the lightcurve on Nov\,16/17 represents a
heating event we 
make use of hardness ratios. Hardness ratios are defined as follows:
\begin{equation}
HR = \frac{B1-B2}{B1+B2}
\label{eq:hrs}
\end{equation}
where $B1$ and $B2$ denote the count rates in a hard and a soft 
band, respectively. 
We defined two hardness ratios for the spectral range of ACIS: 
$HR1$ based on the range from $0.2-1.0$\,keV as the
soft band, and $1.0-2.0$\,keV as the hard band. And $HR2$ with 
$1.0-2.0$\,keV representing the soft band and $2.0-8.0$\,keV
representing the hard band. 
In Fig.~\ref{fig:hrs} these hardness ratios are plotted for the four time segments
introduced above. A hardening of the spectrum with a significance 
of $\approx2\,\sigma$ is apparent 
during the second part of the Nov\,16/17 observation. This may be seen as an 
indication that indeed V410\,Tau underwent a flare, because flares are a
result of (magnetic) heating. 

\section{Discussion}\label{sect:discussion}

We have carried out an intensive
coordinated monitoring campaign in the optical and X-ray wavelength ranges
with the aim to study correlations between the
photometric rotation cycle of V410\,Tau and different activity diagnostics.  
Combining our new data with historic photometric measurements  
we re-examined the variability in the optical lightcurve of V410\,Tau
on various timescales: (i) spot (rotation) cycle, and (ii) long-term (activity)
cycle.

\subsection{Variability in the spot rotation cycle}\label{subsect:disc_spot}

Optical photometric observations were performed at three sites around
the globe, thus providing complete phase coverage of the 1.87\,d spot
cycle despite the short monitoring time of 11\,d. This has allowed us
to measure the time of the minimum in the $V$ band lightcurve of V410\,Tau 
with high precision. 
Combining this measurement with lightcurves from the years 1990 to 2001 
we derived an update of the rotational period. The need for a revision
of the period was indicated by a systematic shift of the times of minimum
in seasonally averaged optical lightcurves folded with the most recent
ephemeris for V410\,Tau given by P94.  
We attribute this to the fact that P94 based their period determination on data
acquired between 1986 and 1992, where the time of minimum seems to have
varied erratically.
The new value for the period is slightly smaller than the period given by P94. 

An alternative explanation for the monotonically increasing shift of the
time of minimum light observed over the last 10\,yrs is latitudinal
migration of spots on the differentially rotating surface. 
However, making use of the differential rotation parameter of V410\,Tau  
known from Doppler imaging we find that the associated change in period would 
correspond to a quite large movement of the spot on the star, which is not 
supported by the photometry and Doppler images. 
Therefore we consider this scenario unlikely as an explanation for the 
systematic effect observed since 1990. Nevertheless, spot migration may be 
responsible for the irregular shifts of the time of minimum observed in the 
years before.         
Unfortunately, no Doppler images are available for these early years.  
But spot models adapted to the photometric data imply indeed changes
in the surface distribution (latitude, longitude, and filling factor) of
the spots (\cite{Herbst89.1}, \cite{Bouvier89.1}). 

In our series of high-resolution spectra we detected RV 
variations which can be explained by the distortions that a spot induces 
onto the absorption line profiles of a rotating star. 
In our RV curve
we included data obtained in 1993 next to our recent monitoring from Nov 2001.
The RV measurements of these two years agree very well in both phase 
(if folded with the new period) and amplitude
demonstrating that the spot distribution has not changed significantly 
over the last decade. 
The half-amplitude of the measured variability is $\Delta V_{\rm rad} \sim 9$\,km/s. 
The RV curve of V410\,Tau was also measured by \citey{Welty95.1}.
Their data showed an indication for an increase in amplitude from 
$\sim 6$ to $\sim 8$\,km/s from 1992 to 1993, while the overall shape remained
stable. These small amplitude variations are consistent with the changes 
in the $V$ band photometry of the same years, and might be related to longterm
changes in the spot size. 

In recent years detailed RV studies with high-precision have been carried out 
for mostly solar-type stars with the aim of detecting low-mass (planetary) companions 
around them (see e.g. \cite{Queloz01.1} for a review). 
However, in active stars spot-related variability may dominate the RV time series,
such that the detection of planets is impeded. Therefore, it is important to assess
the amount of perturbation induced by star spots. 
\citey{Saar97.1} calculated the maximum perturbation of the RV that a 
spot distribution with a filling factor $f_{\rm s}$ 
induces onto a star: $A_{\rm s} \approx 6.5 \times f_{\rm s}^{0.9} \times v\sin{i}$. 
By their definition $f_{\rm s}$ characterizes the inhomogeneous 
part of the spot pattern. 
If we apply their Eq.~1 (given above) to V410\,Tau we derive 
$f_{\rm s} \sim 29$\,\%,  
an almost perfect match to the actual size of the spot on V410\,Tau 
inferred from Doppler imaging ($\sim 30$\,\%; \cite{Joncour94.1}).
Our estimate represents the first attempt to extend the Saar-Donahue relation
to fast rotating (and very active) stars, and we conclude that this relation seems
to hold also in this regime.  
Furthermore, our result implies that the spot on V410\,Tau is significantly 
non-uniform. Indeed, Doppler images indicate that the major spot is near the pole
but not quite identical with a `polar cap'. 

There seems to be no other component in the RV data besides the spot induced variability.
We use the uncertainties in the individual RV measurements ($\sim 3$\,km/s) 
to determine an upper limit for the mass of a short-period 
binary companion possibly hidden in the data. Assuming a circular orbit for the 
hypothesized companion, we can exclude such objects with a mass of 
$m\,sin\,i > 0.03\,M_\odot$ and a period of 
$10$\,d or less. Due to the short timescale of our monitoring no information can be
obtained about longer period companions.

\subsection{Detection of an activity cycle ?}\label{subsect:disc_cycle}

Inspection of the $V$ band lightcurves of V410\,Tau observed during the 
last two decades 
indicates that its behavior has become systematically more stable
and its variability more regular: 
The first photometric data set which provides precise information on both the
observing time and the magnitude is from 1981. 
Around this time a distinct double-peaked
structure evolved, which smoothly transformed itself into a single-peaked 
lightcurve 
over the following five years. This change went along with an increase in the
amplitude by a factor of $2-3$. Since then the simple near-to sinusoidal 
shape of the lightcurve has persisted, only subject to comparatively 
small variations. The most obvious variation within the last ten years 
is the systematic pattern seen in the mean amplitude. 
From a simple harmonic fit to the seasonal averages of the $V$ band magnitude 
of the years 1990 to 2001 
a periodic variation of $5.4$\,yr length is suggested that might represent an 
activity cycle.  

\citey{Baliunas96.1} have outlined a connection between the ratio of cycle
and rotation periods, $P_{\rm cyc}/P_{\rm rot}$, with the dynamo number, $D$, measuring the
efficiency of the stellar dynamo. According to stellar dynamo theory 
$D$ is directly proportional to 
$1/P_{\rm rot}$. From observations of chromospherically active, slowly rotating
stars they found that the empirical slope in the  $\lg{(P_{\rm cyc}/P_{\rm rot})}$ versus 
$\lg{(1/P_{\rm rot})}$ diagram is $\approx 0.74$. 
Interpreting the $5.4$\,yr-variability of V410\,Tau as a cycle period we find that 
its location in the $\lg{(P_{\rm cyc}/P_{\rm rot})} - \lg{(1/P_{\rm rot})}-$diagram
is consistent with the extrapolation of the line identified by 
\citey{Baliunas96.1} into the regime of fast rotating stars. This result is remarkable 
because it seems to indicate that the dynamo process on PMS stars is very similar 
to that on more evolved stars. 
To the best of our knowledge V410\,Tau is the first PMS star investigated in this respect. 

To extend the comparison with the evolved stars we examined the
position of V410\,Tau in the 
$\lg{(\omega_{\rm cyc}/\Omega_{\rm rot})} - \lg{R_{\rm 0}^{-1}}-$diagram, where 
$R_{\rm 0} = 2 \tau_{\rm conv} \Omega_{\rm rot}$ is the Rossby-number, and $\tau_{\rm conv}$
the convective turnover time. 
\citey{Saar99.1} derived $\tau_{\rm conv}$ for their sample of chromospherically active
stars (mostly objects from the Mt.Wilson Ca\,II H+K survey) 
from the models of \citey{Gunn98.1}, 
and identified three branches in this plot: inactive, active, and
superactive stars. 
To estimate $\tau_{\rm conv}$ for V410\,Tau we compare its stellar parameters to 
the PMS calculations by \citey{Ventura98.1}, 
and find $\lg{\tau_{\rm conv}}\,{\rm [d]} \approx 2.3$.  
Because we have found agreement within $0.2$\,dex between zero-age main sequence
convective turnover times estimated from models by \citey{Ventura98.1} 
and those from  models by \citey{Gunn98.1}, we can now make valid comparisons
between V410\,Tau and main-sequence stars. 
We find that the location of V410\,Tau is not consistent with any of
the regions defined by the evolved stars in the 
$\lg{(\omega_{\rm cyc}/\Omega_{\rm rot})} - \lg{R_{\rm 0}^{-1}}-$diagram. 
This is presumably due to its fast rotation combined with a 
large convective turnover time, which is a consequence of its PMS nature. 
Indeed, most PMS stars -- typically characterized by $P_{\rm rot} \leq 10$\,d and
$\tau_{\rm conv} \sim 60 .... 400$\,d -- are expected to lie to the right of the 
active/inactive
branches in this diagram ($\lg{R_{\rm 0}^{-1}} \geq 2$), and below these branches.
Given their generally fast rotation only very short cycle periods
would place them at the extension of these lines. 

V410\,Tau is younger ($\sim 1$\,Myr) 
than all other stars for which activity cycles have been reported so far,
representing a unique test case for dynamo action on the PMS.
Among the stars that seem to display cyclic behavior it comes closest 
to the young solar-analogs, such as AB\,Dor, EK\,Dra, and LQ\,Hya, 
which are single stars characterized by an age of $\sim 50 - 80$\,Myr, 
spectral type of early K, and fast rotation ($P_{\rm rot} < 3$\,d). 
A cycle length for AB\,Dor of $5.3$\,yr was given by \citey{Amado01.1}. 
\citey{Berdyugina02.1} identified three cycles of $5.2$, $7.7$, and $15$\,yr
duration in LQ\,Hya. The photometry of EK\,Dra indicates longterm fading
over the last 35 years (\cite{Froehlich02.1}), 
while \citey{Saar99.1} have predicted cycle periods
of $1.4$ and $39$\,yr for this star. 

The observation of activity cycles in young stars is of paramount importance 
for the understanding of the dynamo operating in these objects. The solar-type
$\alpha\Omega$-dynamo is thought to be localized in the overshoot layer at
the bottom of the convection zone. But in stars with deep convective
envelope a `distributed' dynamo located throughout the convective layer may take
over. 
Furthermore, both observations and theory agree in that differential rotation
is suppressed in fast rotating young stars (\cite{Henry95.1}, \cite{Kueker97.1}).
Therefore, the $\alpha$-effect should dominate over the rotational shear, and a pure
$\alpha\Omega$-dynamo is unlikely to hold. 
The type and stability of the dynamo solutions depend critically on
the strength of differential rotation. For small values of the differential
rotation non-axisymmetric modes are preferred (\cite{Moss95.1}). 
These modes seem not to oscillate (\cite{Kueker99.1}) and are 
difficult to reconcile with the possible observation of an activity cycle
on V410\,Tau. 
In fact, the apparent absence of such cycles in PMS stars is usually taken as
evidence for the action of a non-solar dynamo. 

However, \citey{Kitchatinov01.1} have shown that a transition of the
dynamo mode takes place at an age of $\sim 5-10$\,Myr, such that in the older stars
an axisymmetric oscillating field is preferred. The age of V410\,Tau is 
close to this critical range, 
indicating that it may be a transition object. This might explain the presence
of an activity cycle (representative for oscillating fields) in conjunction 
with long-lived spots (indicating very stable field structures). 

Our present knowledge draws a complicated picture for the magnetic activity 
of V410\,Tau, and PMS stars in general. 
We stress that our conclusions rely on the observation of just one tentative 
cycle period. 
The systematic pattern in the seasonally averaged mean magnitude seems
to have its onset a few years before 1990, at the same time when the single-peaked
lightcurve developed. Earlier than $\sim$ 1986 the lightcurve shows no signs
for cyclic variability. Indeed, in years of low amplitude the $V$ band magnitude
assumes an intermediate level, while a value near maximum would be expected
if a decrease in spot size and number during a cycle minimum were responsible.

\subsection{Spot Rotation Cycle and Activity Diagnostics}\label{subsect:disc_act}

Simultaneously with the optical photometric lightcurve we have acquired
{\em Chandra} X-ray observations and optical spectroscopy. 
{\em Chandra} has targeted V410\,Tau twice during our campaign, at minimum and
maximum optical brightness, respectively. 
Different count levels were found, but their relation to the rotation cycle 
was not confirmed by a following measurement carried out some months later 
that showed an intermediate count rate although obtained at the same
phase as one of the observations from Nov 2001. A seeming lack 
of a correlation between X-ray and optical emission was already pointed 
out by our analysis of archived {\em ROSAT} data (\cite{Stelzer02.1}): we
found that the count rates varied from one observation to the other, however,
without clear relation to the rotation cycle.  
Similarly the H$\alpha$ equivalent width seems not to show a trend related
to the rotational phase. 
Unfortunately, we were not able to obtain full phase
coverage in the optical spectroscopy due to poor weather conditions. 

Reports on coordinated multi-wavelength monitoring of PMS stars are 
scarce in the literature. 
Multi-wavelength observations of TTS in the Taurus star forming region using 
{\em ROSAT} jointly with optical telescopes have revealed flares in Balmer lines
and in X-rays (\cite{Guenther00.1}). 
But due to unfortunate conditions during none of these events was observed  
simultaneously in both the optical and the X-ray range. 
In the same study a weak correlation 
between the X-ray and H$\alpha$ emission was seen for the wTTS V773\,Tau,
suggestive of a relation between the emission sites.  
Simultaneous X-ray and optical observations of 
BP\,Tau were discussed by \citey{Gullbring97.1}. 
No signs for any correlation between the optical and the X-ray emission was seen:
Two optical flares had no counterpart in X-rays, and the data set did not 
allow them to examine variations related to the rotation of the star.  
However, BP\,Tau is a cTTS, such that variability (in both optical and X-rays) 
may be induced by accretion. Therefore, it may not be directly comparable to 
V410\,Tau which is a non-accreting wTTS where all variability should be
related to magnetic activity similar to more evolved stars. 

Recently, \citey{Jardine02.1} have modeled the X-ray emission of AB\,Dor,
for which \citey{Kuerster97.1} did not find any evidence for rotational modulation
of the X-ray emission during monitoring with {\em ROSAT}. 
The model of \citey{Jardine02.1} is based on Zeeman Doppler maps of the surface 
magnetic field structure, and predicts 
little rotational modulation of the X-ray emission due to the extended structure 
and/or high latitudes of coronal features. AB\,Dor is characterized by dark spots
at all latitudes (\cite{Donati97.1}). On V410\,Tau the dominant spot seems to be 
located near the pole, although not completely symmetrically 
(\cite{Hatzes95.1}). But the absence of rotational modulation
in the X-ray emission of V410\,Tau is consistent with a symmetric distribution
of X-ray flux with respect to the rotation axis. Therefore, the spot may not be
the main site of X-ray production.

\subsection{Relation between X-ray and H$\alpha$ flux}\label{subsect:disc_lx_lha}

Coronal X-rays and chromospheric H$\alpha$ emission are thought to be produced
by the same magnetic heating mechanisms which are thought to involve either 
magnetic waves (\cite{Goossens94.1}) or magnetic reconnection processes 
(\cite{Priest00.1}) as heating agent.
Therefore it is expected to find a correlation between chromospheric and
coronal emission 
when comparing stars at different activity levels. In fact \citey{Fleming88.1} 
and \citey{Doyle89.1} have found such a correlation for a sample of dMe flare stars. 
For the PMS no comparable studies exist mainly because of a lack of H$\alpha$ flux
measurements. As we will show in a subsequent paper (Fern\'andez et al., in prep)
the non-periodic component of the photometric variability of 
V410\,Tau is reminiscent of flare stars, 
suggesting that a comparison of its activity to that latter class of objects is
justified. 

\begin{table}
\begin{center}
\caption{X-ray and H$\alpha$ flux and respective luminosities of V410\,Tau during the campaign in November 2001. X-ray emission was measured in the $0.4-8$\,keV energy interval.}
\label{tab:lx_lha}
\begin{tabular}{lcccc} \hline
 {\em Chandra} &                   &         & $f_{\rm x}$  & $\lg{L_{\rm x}}$ \\
 Obs-ID        &                   &         & [${\rm erg\,cm^{-2}\,s^{-1}}$] & [erg/s] \\
\hline
200130Q &                   &         & $1.3 \times 10^{-12}$ & $30.5$ \\
200190  &                   &         & $1.8 \times 10^{-12}$ & $30.6$ \\
200191  &                   &         & $1.6 \times 10^{-12}$ & $30.5$ \\
\hline
        & $W_{\rm H\alpha}$ & $R_{\rm c}$ & $f_{\rm H\alpha}$  & $\lg{L_{\rm H\alpha}}$ \\
        & [\AA]             & [mag]   & [${\rm erg\,cm^{-2}\,s^{-1}}$] & [erg/s] \\
\hline
Min & $1.15$            & $10.35$ & $1.9 \times 10^{-13}$ & $29.6$ \\
Max & $2.45$            & $ 9.96$ & $5.7 \times 10^{-13}$ & $30.1$ \\
\hline
\end{tabular}
\end{center}
\end{table}

We estimated the X-ray flux of V410\,Tau from the ACIS spectrum.
The result is given in Table~\ref{tab:lx_lha}.
Our optical spectra are not flux-calibrated. But the H$\alpha$ flux can
be calculated from $W_{\rm H\alpha}$ with help of the photometry that we
can use to determine the continuum flux, because of the very regular
pattern of the lightcurve.
The $W_{\rm H\alpha}$ was measured on the low-resolution spectra, taking
into account the photospheric absorption. 
We multiplied the $W_{\rm H\alpha}$ by the
specific flux of the R$_{\rm c}$ band (in erg s$^{-1} $cm$^{-2}$ \AA$^{-1}$)
and used the {\em Hipparcos} distance for
V410\,Tau of $136$\,pc (\cite{Wichmann98.1}) to compute $L_{\rm H\alpha}$.
The specific flux for a star with $R_{\rm c}=0$\,mag can be obtained from
\citey{Rydgren84.1}, taking
into account the central wavelength of the band. 
We compute the H$\alpha$ flux for the
minimum and maximum equivalent width
measured in the low-resolution spectra during the
quiescence of the star.
The corresponding $R$ band brightness is extracted from the
lightcurve at the same rotational phase.
 The minimum and maximum quiescent H$\alpha$ flux during our campaign
derived in this way is tabulated in Table~\ref{tab:lx_lha}.

For the M dwarfs studied 
the X-ray luminosity seems to be somewhat higher than the H$\alpha$
luminosity. \citey{Fleming88.1} found the relation $L_{\rm H\alpha} \sim 0.7 L_{\rm x}$.  
Similarly, \citey{Hawley96.1} quoted values between $\sim 0.2 ... 1$ for 
$\lg{(L_{\rm x}/L_{\rm H\alpha})}$ for both field M dwarfs and the zero-age main sequence
cluster IC\,2602. 
The values we derive for the X-ray and H$\alpha$ luminosity of 
V410\,Tau are in good agreement with these relations, suggesting a
tight connection between the activity of this PMS star and that of MS flare stars.

\section{Conclusion}\label{sect:conclusion}

We have presented the results from a coordinated multi-wavelength observing campaign 
for the wTTS V410\,Tau 
aiming to disentangle the role that the various atmospheric layers play in magnetic 
activity.
A multi-wavelength approach is essential in establishing the structure and relation
between the emission sites. 

Our observations of V410\,Tau can be summarized as follows: 
\begin{itemize}
\item An update of the period and ephemeris of the $1.87$\,d cycle 
representing the rotation period of the spot dominating since $\sim$ 1990. 
\item The non-detection of rotational modulation in the X-ray and 
H$\alpha$ lightcurves suggests
either very long coronal loops or that X-ray emission is not concentrated in the
major spot, which extends over the pole of the star but not in a symmetrical way.
\item The first tentative detection of an activity cycle on a PMS star seems to indicate
that the field generation mechanism that drives activity in young stars 
may be more similar to the standard solar-type dynamo than widely believed. 
\item The detection of flares in both the optical and the X-ray regime, as well as the
observed relation between X-ray and H$\alpha$ flux put V410\,Tau in close vicinity 
to the more evolved dMe flare stars.  
\end{itemize}
Whether the characteristics of V410\,Tau are peculiar or typical for the PMS,
or whether it represents a rare kind of a transition object remains open to date. 
We would like to encourage systematic multi-wavelength studies of other young stars 
that will eventually put the nature of magnetic activity on the PMS in a clear 
context to the case of the Sun and solar-like stars.

\begin{acknowledgements}
BS acknowledges financial support from the European Union by the Marie
Curie Fellowship Contract No. HPMD-CT-2000-00013.
MF is partially supported by the Spanish grant PB97-1438-C02-02. 
JFG and VC were supported by grant
POCTI/1999/FIS/34549 approved by FCT and POCTI, with funds from the
European Community program FEDER. 
We thank the referee W. Herbst for constructive comments. 
We also want to acknowlege W.Herbst for maintaining the T Tauri
photometric database (at http://www.astro.wesleyan.edu/~bill/). 
Finally, we thank P. Amado for careful reading of the manuscript. 
This research is partly based on data obtained at the $90$\,cm and $1.5$\,m
telescopes of the Sierra Nevada Observatory (operated by the
Consejo Superior de Investigaciones Cient\'{\i}ficas through the Instituto de
Astrof\'{\i}sica de Andaluc\'{\i}a), the German-Spanish Astronomical Centre on  
Calar Alto (operated by the Max-Planck-Institut f\"ur Astronomie, Heidelberg  
jointly with the Spanish National Commission for Astronomy), and the Lick
Observatory (operated by the University of California).
\end{acknowledgements}

\end{document}